\documentclass[12pt]{article}
\usepackage{amssymb,amsmath,amsthm,amsxtra,amsfonts}
\usepackage{geometry}
\usepackage{aliascnt}
\usepackage{color}
\usepackage[usenames,dvipsnames,table]{xcolor}
\usepackage{subcaption}
\usepackage{bbm}
\usepackage{dsfont}
\usepackage{lscape}
\usepackage{graphicx}
\usepackage{mathrsfs}
\usepackage{tikz}
\usetikzlibrary{plotmarks,arrows,calc,patterns}
\usepackage[round]{natbib}
\usepackage[colorlinks=true,citecolor=Blue,urlcolor=Blue,linkcolor=Blue]{hyperref}
\usepackage{algorithm,algpseudocode}
\usepackage{threeparttable}
\usepackage{booktabs}
\usepackage{rotating}
\usepackage{algorithm}
\definecolor{warm-green}{rgb}{0.098, 0.745, 0.290}

\newtheorem{example}{Example}[section]

\newtheorem*{assumption*}{Assumption}

\newcommand{\R}{\ensuremath{\mathbf{R}}}
\newcommand{\Z}{\ensuremath{\mathbf{Z}}}

\newcommand{\Nn}{\ensuremath{\mathbf{N}}}

\newcommand{\E}{\ensuremath{\mathbb{E}}}

\newcommand{\Var}{\ensuremath{\mathrm{Var}}}

\DeclareMathOperator*{\argmin}{arg\,min}

\newcommand{\one}{\ensuremath{\mathds{1}}}

\begin{document}
	\title{Econometrics of Machine Learning Methods in Economic Forecasting}
	
	\author{Andrii Babii\thanks{University of North Carolina at Chapel Hill - Gardner Hall, CB 3305 Chapel Hill, NC 27599-3305. Email: babii.andrii@gmail.com.} \and Eric Ghysels\thanks{Department of Economics and Kenan-Flagler Business School, University of North Carolina--Chapel Hill. Email: eghysels@unc.edu.} \and Jonas Striaukas\thanks{{Department of Finance, Copenhagen Business School, Frederiksberg, Denmark. Email: jonas.striaukas@gmail.com.}}}
	
	\maketitle

	\vfill
\thispagestyle{empty}

\setcounter{page}{0}

\newpage

%\paragraph{Background}
%The emphasis of the Handbooks is on the practical application of research methods to both quantitative and qualitative evidence. The idea is that the Handbook will assist practising researchers in generating robust research findings that policymakers can use with confidence. 

\section{Introduction}
Economic forecasting has traditionally relied on simple models estimated with the maximum likelihood (MLE) approach.  The limitations of the MLE are well known as eloquently described in \cite{bradley2021computer}:

\smallskip

{\footnotesize ``Arguably the 20th century’s most influential piece of applied mathematics, maximum likelihood continues to be a prime method of choice in the statistician’s toolkit. Roughly speaking, maximum likelihood provides nearly unbiased estimates of nearly minimum variance, and does so in an automatic way.	That being said, maximum likelihood estimation has shown itself to be an inadequate and dangerous tool in many 21st century applications. Again speaking roughly, unbiased can be an unavoidable luxury when there are hundreds or thousands of parameters to estimate at the same time."}

\smallskip

\cite{james1961estimation} made this point dramatically in a much simpler setting involving just a couple of parameters. The machine learning (ML) methods developed over roughly the past 60 years have revolutionized decision-making across various fields. At its core, ML involves formulating a loss or cost function for forecasting rules. In this context, a forecasting rule, denoted as $f(x_t)$, predicts the value of a target variable, $y_{t+h}$, at a future horizon, $h$, based on information available at time $t$. The loss function, $\ell(y_{t+h},f(x_t))$, quantifies the error incurred by the forecasted value compared to the actual outcome.

\smallskip

The central goal is to approximate the optimal decision rule, $f^*$, which minimizes the expected loss, $\E[\ell(y_{t+h},f(x_t))]$. This approach has its roots in the decision theory, see \cite{wald1949statistical}, and is adopted in statistical learning, see \cite{vapnik1999nature}, and economic forecasting, see \cite{granger2000economic}. For instance, when employing a quadratic loss function, $\ell(y_{t+h},f(x_t)) = (y_{t+h} - f(x_t))^2$, the optimal decision rule corresponds to the (non-linear) regression, $f^*(x_t)=\E[y_{t+h}|x_t]$ with respect to $f(x_t)$.\footnote{Equivalently, we could consider the regression model, $y_{t+h} = f^*(x_t) + \varepsilon_{t+h}$ with $ \E[\varepsilon_{t+h}|x_t]=0$.}

\smallskip

The data-driven decision rules lead to the bias-variance trade-off in the forecasting performance. Flexible nonparametric techniques offer a solution by reducing bias at the cost of increasing variance, leading to potential overfitting issues. At the same time, regularization and dimensionality reduction introduce some bias to reduce the variance. Machine learning offers a wide array of nonparametric and high-dimensional tools, enabling flexible and accurate approximations of the optimal decision rules, adapting to the bias-variance trade-off, and optimizing the forecasting performance.

\smallskip

Many of the widely used ML tools relate to known and well-established statistical methods. For example, deep learning can be understood as a regression model with nonlinearities generated by a multi-layer neural network; see \cite{hornik1990universal} and \cite{chen2007large}.\footnote{Various forms of neural networks have achieved a remarkable performance recently with perceptual data like text, images, speech, or videos; see also \cite{farrell2021deep} and \cite{gu2020empirical} for applications with tabular data.} Random forests and gradient boosting which can be understood as a new generation of regression and classification trees; see \cite{breiman1984classification}. The penalized regression can be traced back to the idea of shrinkage, see \cite{james1961estimation}, regularization of ill-posed inverse problems, see \cite{tikhonov1963solution}, and the ridge regression, see \cite{hoerl1970ridgeb,hoerl1970ridge}.\footnote{See also \cite{carrasco2007linear} and \cite{babii2017completeness}.} 

\smallskip

While the development of ML methods has a long history, the remarkable recent success and wide adoption are mostly due to the increasing availability of new high-dimensional data, cheap computational power, and scalable statistical packages.\footnote{One may quote the remarkable success of ML methods in the prediction contests with substantial monetary prizes, such as Kaggle or Makridakis Competitions; see \cite{makridakis2020m4,makridakis2022m5}.} Economists also rely increasingly on high-dimensional datasets such as textual and image data, credit card spending, or Google Trends. Consequently, ML methods are gaining appreciation and are becoming ubiquitous in economics and finance.

\smallskip

In this chapter, we aim to review some of the recent developments in the machine learning literature for economic forecasting, focusing on the appropriate treatment of time series lags, panel and tensor data, nowcasting, high-dimensional Granger causality tests, time series cross-validations, and classification. Hence, this chapter is focused, given the space limitations, on topics of interest to forecasters and we refer the reader to other existing surveys and introductions to the ML methods for a more general review of the subject.\footnote{See \cite{james2013introduction}, \cite{hastie2009elements}, and \cite{breiman2001statistical} for general introductions. See also \cite{mullainathan2017machine}, \cite{athey2019machine}, and \cite{varian2014big} for economics surveys. Lastly, see \cite{coulombe2019machine} and \cite{masini2023machine} for time series reviews.}

\smallskip

\section{High-Dimensional Projections}
\subsection{Time Series Forecasting}
The empirical analysis of time series data entails several notable challenges.  Firstly, in a data-rich time series environment the objective is often to forecast a low-frequency variable (e.g.\ quarterly GDP growth or inflation) while the information set may contain predictors measured at a higher frequency (e.g.\ monthly or daily). Additionally, certain economic effects tend to persist over time. This brings us to the question of how to combine the high (or same) frequency time series lags in regression equations. 

\smallskip

Secondly, the prevalence of high-dimensional datasets further compounds the complexity. In addition to traditional macroeconomic and financial indicators, modern empirical research increasingly relies on non-standard data sources like textual data, credit card spending records, traffic and satellite data, among others. Consequently, the task of selecting an accurate forecasting model from this vast array of predictors becomes a significant challenge. Shrinkage methods like ridge regression or LASSO effectively mitigate multicollinearity and overfitting issues, making them particularly suited for handling large predictor sets.

\smallskip

To address the aforementioned challenges, \cite{babii2021machine} introduce high-dimensional regularized projections for time series data inspired by the mixed-frequency data sampling, i.e.\ MIDAS  regression or the distributed lag econometric literature (see  \cite{ghysels2006predicting}).  Let $(y_t)_{t\in[T]}$  be a target time series, e.g.\ quarterly GDP growth or inflation (where we put $[T]=\{1,2,\dots,T\}$ for a positive integer $T$). The covariates consist of $K$ time-varying predictors measured potentially at higher frequencies, e.g.\ quarterly, monthly, or daily,
\begin{equation*}
	\left\{x_{t-j/n_k^H,k}:\;,t\in[T],j=0,\dots,n^L_kn_k^H-1,k\in[K]\right\},
\end{equation*}
where $n_k^H$ is the number of high-frequency observations for the $k^{\rm th}$ covariate in a low-frequency time $t$, and $n^L_k$ is the number of low-frequency periods used as lags. For instance, $n^L_k=1$ corresponds to a single quarter of high-frequency lags used as covariates and $n_k^H=3$ corresponds to 3 months of data available per quarter.

\smallskip

The mixed frequency time series regression equation for forecasting a low -frequency target $y_{t+h}$ at a horizon $h\geq 1$ is  
\begin{equation*}
	y_{t+h} = \alpha + \sum_{j=0}^J\rho_jy_{t-j}  +  \sum_{k=1}^{K}\psi(L^{1/n_k^H};\beta_k)x_{t,k} + u_{t+h},
\end{equation*}
where we use the lag polynomial notation
	$\psi(L^{1/n_k^H};\beta_k)x_{t,k}$ = $\frac{1}{m_k}\sum_{j=0}^{m_k-1}\beta_{j,k}x_{t-j/n_k^H,k},$
where $m_k=n_k^Ln_k^H$ is the total number of all lags. The resulting projection model has a large number of parameters and is prone to overfitting.

\smallskip

\cite{babii2021machine} propose to parametrize the lag coefficients using a MIDAS weight function $\omega$ described with a low-dimensional parameter $\beta_k\in\R^L$
\begin{equation*}
	\psi(L^{1/n_k^H};\beta_k)x_{t,k} = 
	\frac{1}{m_k}\sum_{j=0}^{m_k-1}\omega\left(\frac{j}{n_k^H};\beta_k\right)x_{t-j/n_k,k},
\end{equation*}
where
$\omega(s;\beta_k)$ = $\sum_{l=0}^{L-1}\beta_{l,k}w_l(s)$  and
$(w_l)_{l\geq 0}$ is a collection of approximating functions, called \textit{dictionary}. The default choice for the dictionary could be a set of Legendre polynomials shifted to $[0,n_k^L]$ interval.\footnote{Other possibilities include splines, trigonometric polynomials, or wavelets.} Given this choice, the forecasting equation is mapped to the linear regression model, where covariates are weighted by a matrix generated from the weight function. Importantly, the time series lags define the sparse-group structure, where a group of coefficients $\beta_k\in\R^L$ corresponding to the $k^{\rm th}$ covariate, is approximately sparse. 

\smallskip

Next, \cite{babii2021machine} propose to use the sparse-group LASSO (sg-LASSO) estimator of  \cite{simon2013sparse}, namely:
\begin{equation*}
	\min_{b\in\R^p}\|\mathbf{y} - \mathbf{X}b \|_{T}^2 +  \lambda\Omega(b),
\end{equation*}
where $\|.\|_T= |.|_2/\sqrt{T}$  is the empirical norm, $\lambda\geq 0$ is a tuning parameter, and $\Omega$ is the sg-LASSO regularizing functional:
\begin{equation*}
	\Omega(b) = \gamma|b|_1 + (1-\gamma)\|b\|_{2,1},
\end{equation*}
is a penalty function.\footnote{We use $|z|_q=(\sum_{i=1}^p z_i^q)^{1/q}$ to denote the $\ell_q$ norm of $z\in\R^p$.} The sg-LASSO penalty is a linear combination of the LASSO ($\ell_1$ norm), see \cite{tibshirani1996regression},  and the group LASSO, see \cite{yuan2006model} ($\|\beta\|_{2,1}=\sum_{k=1}^K|\beta_k|_2$). The group LASSO penalty selects covariates while the standard LASSO penalty selects the shape of the MIDAS weight function.

\smallskip

Hence, the setting nests the standard LASSO ($\gamma=1$) and the group LASSO ($\gamma=0$) as special cases. \cite{babii2021machine} establish the non-asymptotic theoretical properties of the estimator for heavy-tailed $\tau$-mixing processes which are general enough for macroeconomic and financial time series.\footnote{This class of processes is large enough to cover the $\alpha$-mixing processes as well as infinite linear transformations of $\beta$-mixing processes.} The properties rely on the Fuk-Nagaev concentration inequality obtained in \cite{babii2022high}.

\smallskip

The literature on the applications of penalized regressions to time series data is vast and we can only mention some of the interesting developments. \cite{mogliani2021bayesian} propose a Bayesian approach to the high-dimensional MIDAS regressions based on the group LASSO. They find good forecasting performance in forecasting US economic activity. \cite{beyhum2023factor} extend the work of \cite{babii2021machine} proposing a factor augmented sg-LASSO-MIDAS regression. They find that factor augmentation yields improvements in nowcast accuracy during the COVID period. \cite{hecq2023hierarchical} consider the extension of the sparse-group LASSO, called the hierarchical LASSO, where the groups can be arranged on a multi-level tree.

\smallskip

%\cite{schorfheide2021real} extend the analysis of \cite{schorfheide2015real} to nowcasting during COVID period.  

Some of the penalized methods have been used for a long time. For example, the HP filter is essentially the penalized regression; see  \cite{mei2022boosted} and \cite{phillips2021boosting} for recent contributions. \cite{chen2023time} propose a nonparametric estimator of time-varying forecast combination weights and develop corresponding asymptotic theory. They apply the LASSO-type estimator for kernel regression to estimate the forecast combination weights. Application to inflation and unemployment shows the benefits of the method compared to alternative techniques used in forecast combination literature such as Complete Subset Regressions proposed by \cite{elliott2013complete}, partially egalitarian LASSO approach of \cite{diebold2019machine}.
 
\smallskip

There are also a number of applications of penalized regressions to asset pricing; see \cite{gu2020empirical}, \cite{freyberger2020dissecting}, \cite{feng2020taming}, \cite{bryzgalova2015spurious}, and the review paper  \cite{giglio2022factor}.

\smallskip

On the theory side, \cite{kock2016consistent} and \cite{medeiros2016,medeiros2017adaptive} establish the model selection consistency and derive convergence rates for the adaptive LASSO with time series data. \cite{kock2015oracle} and  \cite{masini2022regularized} derive convergence rates for high-dimensional VAR models; see also \cite{wong2020lasso}, \cite{chernozhukov2021lasso}, \cite{adamek2023lasso} for convergence rates of LASSO under $\beta$-mixing, physical dependence, and near-epoch dependence.

\subsection{Panel Data}
It is often the case that the objective is to forecast or nowcast a large number of long time series of size $T$, e.g.\ $N$ regional growth indices or price/earnings ratios for $N$ firms observed at $T$ quarters. In the latter case, the predictors cover the firm-specific accounting and textual data as well as the aggregate macroeconomic and financial indicators. While the time series methods described in the previous section can be applied series-by-series, this approach ignores the cross-sectional variation in the panel. 

\cite{babii2022machine,babii2023panel} focus on the high-dimensional panel data regressions
\begin{equation*}
	y_{i,t+h} = \alpha_i + \sum_{k=1}^{K}\psi(L^{1/n_k^H};\beta_k)x_{i,t,k} + u_{i,t|\tau},\qquad i\in[N],t\in[T]
\end{equation*}
where the index $i$ denotes the cross-sectional dimension, e.g.\ a region or a firm. The corresponding regularized fixed effects estimator solves
\begin{equation*}\label{eq:sgl}
	\min_{(a,b)\in\R^{N+p}}\|\mathbf{y} - Ba - \mathbf{X}b \|_{NT}^2 + 2 \lambda\Omega(b),
\end{equation*}
where $B=I_N\otimes\iota$ and $\iota\in\R^T$ is an ``all ones" vector and the panel data observations are stacked in $(\mathbf{y},\mathbf{X})$. An attractive feature of the estimator is that it captures the heterogeneity of time series intercepts $\alpha_i$. The disadvantage is that estimating $N$ additional parameters can be statistically costly. In some cases, these costs outweigh the benefits and simpler pooled panel data regressions
\begin{equation*}\label{eq:sgl}
	\min_{(a,b)\in\R^{1+p}}\|\mathbf{y} - \iota a - \mathbf{X}b \|_{NT}^2 + 2 \lambda\Omega(b),
\end{equation*}
where the intercepts are $\alpha_1=\dots=\alpha_N=a$.

\smallskip

Some of the recent empirical work using machine learning, panel data, and nowcasting includes \cite{van2023man} (firm earnings), \cite{ghysels2022real} (government earnings and expenditures), \cite{fosten2022panel} (state-level GDP growth). On the methodology side, \cite{carrasco2016sample} consider general regularization based on spectral decomposition covering the ridge regression as a special case. \cite{carvalho2018arco} apply the LASSO to predict controls for causal inference, generalizing the method of synthetic controls of \cite{abadie2010synthetic}.\footnote{The literature on using LASSO to predict the counterfactuals is fast; see \cite{belloni2014inference}, \cite{chernozhukov2018double}, and references therein.}

\subsection{Nowcasting, real-time data flow, and textual data}
The term nowcasting is a contraction of now and forecasting. It is defined as the prediction of the present, the very near future, or the very recent past, using the real-time data flow reflecting the evolving economic conditions and data revisions;  see \cite{banbura2013now} and \cite{giannone2008nowcasting}. Nowcasting a target variable $y_{t}$ at low-frequency (e.g.\ quarterly) often involves vintage data, defined as a sequence of information sets, denoted
\begin{equation*}
	I_{t_r}=\left\{x_{k,\lceil t_r\rceil - j/n_k^H|r}:\; k\in[K_r],j=\underline{j}_{r,k},\dots,n_k^Hn_k^L-1 \right\}
\end{equation*}
where $t_1\leq t_2\leq \dots\leq t_R$ are times when the information set is updated.\footnote{For a real number $a$, we use $\lceil a\rceil$ to denote the smallest integer larger than $a$.} The updates at a given time $t_r$ appear for two reasons: 1) new data is \textit{released}; 2) old data is \textit{revised}. The revisions are especially common for the macroeconomic data and it is crucial to forecast using the vintage data available at a particular point of time to avoid the look ahead biases (see \cite{ghysels2018forecasting} for further discussion).

\smallskip

\cite{babii2021machine} consider the problem of nowcasting the quarterly US GDP growth using higher frequency macroeconomic and financial. They find that the machine learning nowcasts are either superior or at par with those posted by the New York Federal Reserve Bank. Additional gains are achieved using the data coming from the textual analysis of economic news; see \cite{bybee2020structure}. \cite{ellingsen2022news} also report that the textual news data add value to the traditionally used FRED-MD data. In a related work, \cite{babii2022machine,babii2023panel} consider the problem of nowcasting the price-earnings ratios with firm-specific accounting information as well as the aggregate macroeconomic, financial, and textual news information and report. They find that the machine learning panel data models perform favorably, cf.\ \cite{ball2018automated}.

\smallskip

\cite{borup2023mixed} study the weekly unemployment insurance initial claims using unrestricted MIDAS specification utilizing Google trends data; see also  \cite{ferrara2022google}. They find that the ensemble (or combinations) of linear and nonlinear ML methods perform the best and that the daily Google Trends data were particularly relevant during the COVID-19 crisis.  \cite{jardet2022nowcasting} also find that nowcasting performance improves during crises period. On the other hand, \cite{jardet2022nowcasting} find that nowcasting performance improves during crises period when weekly data is used in prediction models.

\smallskip

\cite{barbaglia2023forecasting} develop a Fine-Grained Aspect-based Sentiment analysis method to compute sentiments from news articles about the state of the economy. They find that economic news sentiments extracted from a large pool of news articles track economic cycles and help accurately nowcast economic activity. Lastly, \cite{cimadomo2021nowcasting} consider large Bayesian Vector Autoregressive (BVAR) models to nowcast the US economic activity.

\subsection{Granger Causality Tests}
The time series models are often misspecified. In this case, the regression has only a projection interpretation and the regression errors are serially correlated. In addition to that the LASSO estimator has a complicated sampling distribution due to a significant shrinkage bias. Let $\hat\beta_G=(\hat \beta_j)_{j\in G}$ be a subset of projection coefficients fitted with the LASSO (or sg-LASSO) indexed by $G\subset[p]$, where $p$ is the number of regressors. \cite{babii2022high}  show that for the heavy-tailed $\tau$-mixing time series, we have
\begin{equation*}
	\sqrt{T}(\hat\beta_G + B_G - \beta_G) \xrightarrow{d} N(0,\Xi_G),
\end{equation*}
where $B_G$ is a bias correction term, $\Xi_G = \lim_{T\to\infty}\Var\left(\frac{1}{\sqrt{T}}\sum_{t=1}^T u_t\Theta_Gx_t\right)$ is the long-run variance, and $\Theta$ is the precision matrix.\footnote{See also \cite{chernozhukov2021lasso} for the physically dependent processes and \cite{adamek2023lasso} for the near-epoch dependent processes.}

\smallskip

The long-run variance can be estimated using the standard HAC estimator, see \cite{newey1987simple} and \cite{andrews1991heteroskedasticity}
\begin{equation*}
	\label{eq:hacformula}
	\hat\Xi_G \triangleq \sum_{|k|<T}K\left(\frac{k}{M_T}\right)\hat{\Gamma}_k,
\end{equation*}
where $K:\R\to[-1,1]$ is the kernel weight function, $M_T$ is the lag truncation parameter, and $\hat{\Gamma}_k = \hat\Theta_G\left(\frac{1}{T}\sum_{t=1}^{T-k}\hat u_t\hat u_{t+k} x_tx_{t+k}^\top\right)\hat\Theta_G^\top$, are the autocovariances for fitted residuals $\hat u_t$. \cite{babii2022high} characterize the MSE convergence rate of the HAC estimator based on the sg-LASSO residuals. Their result leads to the following ``rule of thumb" choice of the bandwidth parameter
\begin{equation*}\label{eq:bw_rules}
	M_T = \begin{cases}
		1.3\left(\frac{T}{\log p}\right)^\frac{1}{1+\varsigma}, & \text{sub-Gaussian data} \\
		1.3 \left(\frac{T^{2-2/q}}{p^{2/q}}\right)^\frac{1}{1+\varsigma}, & \text{heavy-tailed data},
	\end{cases}
\end{equation*}
where $\varsigma=2$ for the Quadratic spectral and Parzen kernels, and $q>2$ is the number of finite moments in the data.

\smallskip

For forecasting problems, we can use these results to test Granger causality which is a formal statistical way to evaluate whether a particular time series marginally adds to the projection of a target variable on a set of predictors. Interestingly, in his original paper, \cite{granger1969investigating} defined causality in terms of high-dimensional time series data which he referred to as ``all the information available in the universe at time $t$". To test whether a series $(w_t)_{t\in\Z}$ Granger causes another series $(y_{t})_{t\in\Z}$ at a horizon $h$, consider
\begin{equation*}\label{eq:granger}
	y_{t+h} = c + \sum_{j\geq 1}z_{t,j}\gamma_j + \mathbf{w}_{t-1}^\top\alpha + u_t,
\end{equation*}
where $(z_{t,j})_{j\geq 1}$ is a high-dimensional set of controls $\alpha\in\R^K$ and $\mathbf{w}_{t-1}\in\R^K$ is a vector lags of $(w_t)_{t\in\Z}$. 

\smallskip

\cite{babii2022high} show that under the null hypothesis, the bias-corrected Wald statistics follows a chi-squared distribution
\begin{equation*}\label{eq:wald}
	W_T := T\left[(\hat\alpha + A - \alpha)\right]^\top \hat\Xi_\alpha^{-1}\left[(\hat\alpha + A - \alpha)\right] \xrightarrow{d} \chi^2_K,
\end{equation*}
where $A$ is the bias correction term and $\hat\Xi_\alpha$ is the HAC estimator.
The practical implementation of the test is as follows:
\begin{enumerate}
	\item Estimate $\alpha\in\R^K$ using the LASSO (or sg-LASSO) and compute the HAC estimator using the LASSO residuals.
	\item Compute the bias-corrected Wald statistics.
	\item Reject the Granger non-causality if $W_T>q_{1-\alpha}$, where $q_{1-\alpha}$ is the $1-\alpha$ quantile of $\chi^2_K$ and do not reject otherwise.
\end{enumerate}

\smallskip

One could alternatively, use a likelihood ratio test or an LM test; see \cite{hecq2023granger} for the latter.

\subsection{Time Series Cross-Validation}
The practical implementation of ML methods requires specifying one or several tuning parameters. For i.i.d.\ data, a common practice is to rely on the $K$-fold cross-validation. which may or may not be appropriate for time series data. \cite{bergmeir2018note} show that for autoregressive models with i.i.d.\ errors the standard $K$-fold cross-validation remains valid. However, with correlated errors - for example, when the regression has only projection interpretation due to misspecification - the standard cross-validation fails due to the correlation between the training and the test samples.

\smallskip

One could rely on the following leave-one-out cross-validation with a gap procedure that decorrelates the training and the test samples, see also \cite{chu1991comparison}. Let $\hat f_\lambda(x_t)$ be a prediction rule of a machine learning model with tuning parameters $\lambda=(\lambda_1,\dots,\lambda_M)$.\footnote{For example, the sg-LASSO corresponds to the linear prediction rule $\hat f_\lambda(x_t)=x_t^\top\hat \beta_\lambda$ with $M=2$}. For some $l\in\Nn$ and each $t=1,\dots,T:$
\begin{enumerate}
	\item If $t>l + 1$ and $t < T - l$, use observations $I_{t,l}=\{1,\dots,t-l-1,t+l+1,\dots,T\}$ to fit the machine prediction, denoted $\hat f_{\lambda,-t,l}(x_t)$. For $t=1,\dots,l+1$, use $I_{t,l} = \{t+l+1,\dots, T\}$ as the training sample. Similarly, for $t=T-l,\dots,T$, use $I_{t,l} = \{1,\dots,T-l-1\}$ as the training sample.
	\item Use the left-out observations to test the model
	\begin{equation*}
		CV(\lambda) = \frac{1}{T}\sum_{t=1}^T\ell(y_t-\hat f_{\lambda,-t,l}(x_t)),
	\end{equation*}
	where $\ell$ is the loss function, e.g.\ the MSE or quadratic loss, $\ell(u)=u^2$.
	\item Minimize $CV(\lambda)$ with respect to $\lambda$.
\end{enumerate}

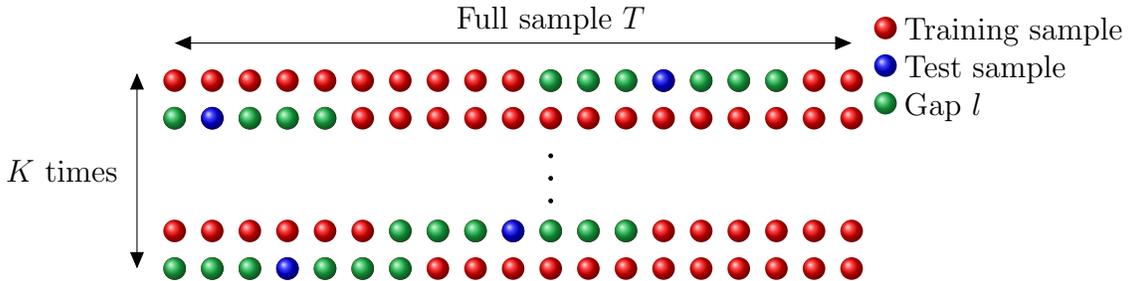
\begin{figure}[h]
	\centering
	\begin{tikzpicture}[shading=ball]
		% legend
		\shade[ball color=warm-green](9.45,2.2) circle (0.15cm); 
		\node[draw=none, text width=4cm] at (11.7,3.15) {Training sample};
		\shade[ball color=blue](9.45,2.7) circle (0.15cm);
		\node[draw=none, text width=4cm] at (11.7,2.65) {Test sample};
		\shade[ball color=red](9.45,3.2) circle (0.15cm);
		\node[draw=none, text width=4cm] at (11.7,2.15) {Gap $l$};
		% full sample text
		\node[draw=none] at (5,3.3) {Full sample $T$};
		\draw[>=triangle 45, <->] (0,3) -- (9.0,3);
		% K times text
		\node[draw=none] at (-1.5,1.3) {$K$ times};
		\draw[>=triangle 45, <->] (-0.5,0) -- (-0.5,2.6);
		% full sample
		\foreach \x in {0.0,0.5,1.0,1.5,2.0,2.5,3.0,3.5,4.0,4.5,5.0,5.5,6.0,6.5,7.0,7.5,8.0,8.5,9.0}
		\foreach \y in {0.0,0.5,2.0,2.5}
		\shade[ball color=red](\x,\y) circle (0.15cm);
		% test set 1
		\foreach \x / \y in {1.5/0.0,
			4.5/0.5,
			0.5/2.0,
			6.5/2.5}
		\shade[ball color=blue](\x,\y) circle (0.15cm);
		% test set 2
		% gap for test set 1
		\foreach \x / \y in {0.0/0.0, 0.5/0.0, 1.0/0.0,  2.0/0.0, 2.5/0.0, 3.0/0.0, 
			3.0/0.5, 3.5/0.5, 4.0/0.5,  5.0/0.5, 5.5/0.5, 6.0/0.5, 
			0.0/2.0,  1.0/2.0, 1.5/2.0, 2.0/2.0,
			5.0/2.5, 5.5/2.5, 6.0/2.5,  7.0/2.5, 7.5/2.5, 8.0/2.5} 
		\shade[ball color=warm-green](\x,\y) circle (0.15cm);
		% gap for test set 2
		% dot dot dot
		\foreach \x / \y in {5.0/0.9, 5.0/1.2, 5.0/1.5}
		\filldraw[color=black](\x,\y) circle (0.025cm);
	\end{tikzpicture}
	\caption{Time series cross-validation scheme with $l$ = 3 \label{fig:cv}}
\end{figure}
For $l=0$ the procedure is the usual leave-one-out cross-validation while of $l\geq 1$, there is a gap of $l$ observations separating the test and the training samples. Since the procedure is computationally demanding, \cite{babii2022high} draw randomly a sub-sample $I\subset [T]$ of size $K$ and minimize
\begin{equation*}
	CV_K(\lambda) = \frac{1}{K}\sum_{t\in I}\ell(y_t - \hat f_{\lambda,-t,l}(x_t))
\end{equation*}
instead. Figure~\ref{fig:cv} illustrates it for a gap of $l=3$ observations. The red training data are separated from the blue test data with a gap of 3 green left-out observations on each side. 

\section{Classification for Economists}
Forecasting binary variables is a prominent problem, also known as the classification or screening in the computer science literature; see \cite{lahiri2013forecasting} for a review. The classification rules build a foundation for the automated data-driven algorithm based on vast data inputs that are increasingly used for various life-changing decisions, including job hiring, pre-trial release from jail, medical testing and treatment. They are also used for various routine tasks such as loan approval, fraud detection, or spam filtering.

\smallskip

\cite{babii2020binary} highlight that the downside risk and upside gains of many economic decisions are not symmetric. The importance of asymmetries in prediction problems arising in economics has been recognized for a long time; see \cite{granger1969prediction}, \cite{manski1989estimation}, \cite{granger2000economic}, \cite{elliott2013predicting}, and the textbook treatment in \cite{graham2016economic}, among many others. At the same time, the standard logistic regression and machine learning algorithm often ignore the asymmetric cost and benefit considerations. Consider, for example, the problem of pre-trial detention where we need to decide on whether to keep an individual in jail before the trial. The decision is $f(x_i)\in\{-1,1\}$ (1 if detained) outcome is $y_i\in\{1,-1 \}$ (1 if individual becomes a recidivist).

	\begin{table}[h]
		\resizebox{1.1\textwidth}{!}{\begin{minipage}{\textwidth}
				\centering
				\begin{tabular}{c|cc}
					decision $\backslash$ outcome & Recidivist & Non-recidivist \\ \hline
					Detained & $EBD(c)+ECD(d)$ & $\psi_GECD(d)$ \\
					Released & $\psi_GC(z,c)$ & 0 \\
					
				\end{tabular}
				\caption{Asymmetric loss for pre-trial detention}
				\label{table:name}
		\end{minipage} }
	\end{table}

If an individual is correctly detained, we have some economic benefits of detention (EBD) which is a function of the type of committed crime $c$ as well as the costs of detention (ECD) depending on the duration $d$. If we incorrectly release a recidivist, there are costs associated with the recidivism, depending on the type of crime $c$ and some other characteristics $z$. The loss function also incorporates the group-specific weights $\psi_G$ with $G\in\{0,1\}$ to have an opportunity to control the group-specific false positive and false negative mistakes which are related to the fairness issues; see \cite{barocas2017fairness}.

More generally, we can summarize the loss function in four states of the world as a quartet of covariate-driven functions $\ell_{f,y}(x)$ with $f,y\in\{-1,1\}$ encoding the economic cost and benefit considerations.\footnote{Compare this with the standard binary classification (e.g.\ logistic regression) which optimizes the following loss function $\ell(f,y,x) = \one\{f(x)\ne y\}$.} \cite{babii2020binary} show that the economic costs and benefits can be accommodated by reweighing the logistic regression (or ML methods) by the asymmetries of the loss function. For instance, in the case of the logistic regression with a LASSO penalty, it is enough to solve
\begin{equation}\label{eq:convex}
	\hat\theta = \argmin_{\theta\in\R^p}\frac{1}{n}\sum_{i=1}^n\omega(y_i,x_i)\log\left(1 + e^{-y_ix_i^\top \theta}\right) + \lambda|\theta|_1,
\end{equation}
where the individual likelihoods are weighted by $\omega(y_i,x_i):= y_ia(x_i)+b(x_i)$ with
\begin{equation*}
	\begin{aligned}
		a(x) & = \ell_{-1,1}(x)-\ell_{1,1}(x) + \ell_{-1,-1}(x) - \ell_{1,-1}(x), \\
		b(x) & = \ell_{-1,1}(x) - \ell_{1,1}(x) + \ell_{1,-1}(x) - \ell_{-1,-1}(x).
	\end{aligned}
\end{equation*}
The data decision rule is then $\hat f(x_i) = 1$ if $x_i^\top\hat\theta\geq0$ and $\hat f(x_i)=-1$ if $x_i^\top\hat\theta<0$. Note that the problem in Eq.~(\ref{eq:convex}) is a convex optimization problem that can be easily solved using the standard optimization methods. This bypasses the need to solve a non-convex problem using the mixed-integer optimization; see  \cite{elliott2013predicting} and \cite{florios2008exact}.\footnote{See also \cite{pellattbinary} for a PAC-Bayesian perspective.} In addition to the (high-dimensional) logistic regression, the approach of \cite{babii2020binary} can also be applied to suitably reweighted support vector machines (SVM), boosting, and deep learning.\footnote{Most of the popular machine learning packages classification packages in Python and R conveniently allow to specify weights including the XGBoost, see \cite{chen2016xgboost}, deep learning, and SVM. }

\smallskip 

Some recent methodological developments and applications related to classification include \cite{barbaglia2023forecasting}, \cite{kitagawa2021constrained}, and \cite{christensen2020robust}, while the fairness issues are also discussed in \cite{rambachan2020economic} and \cite{viviano2023fair}.

\section{Tensor Factor Models}
The datasets available in modern empirical applications often have a multi-dimensional panel structure. For example, in the regional macroeconomic datasets, $y_{i,j,t}$ is the macroeconomic indicator $i$ for region $j$ measured at time $t$, so the data is the 3-dimensional panel. Another example is the network data, where $y_{i,j,t}$ is the outcome for the nodes $(i,j)$ at time $t$, e.g.\ the exchange rates for a pair of currencies $(i,j)$. In asset pricing, $y_{i,j,t}$ is the excess return of $j^{\rm th}$ quantile sorted on anomaly $i$ at time $t$; see  \cite{lettau2020factors}. Adding the international dimension, we obtain the 4-dimensional panel.

\smallskip

\begin{figure}[!ht]
	\centering
	\includegraphics[scale=0.4]{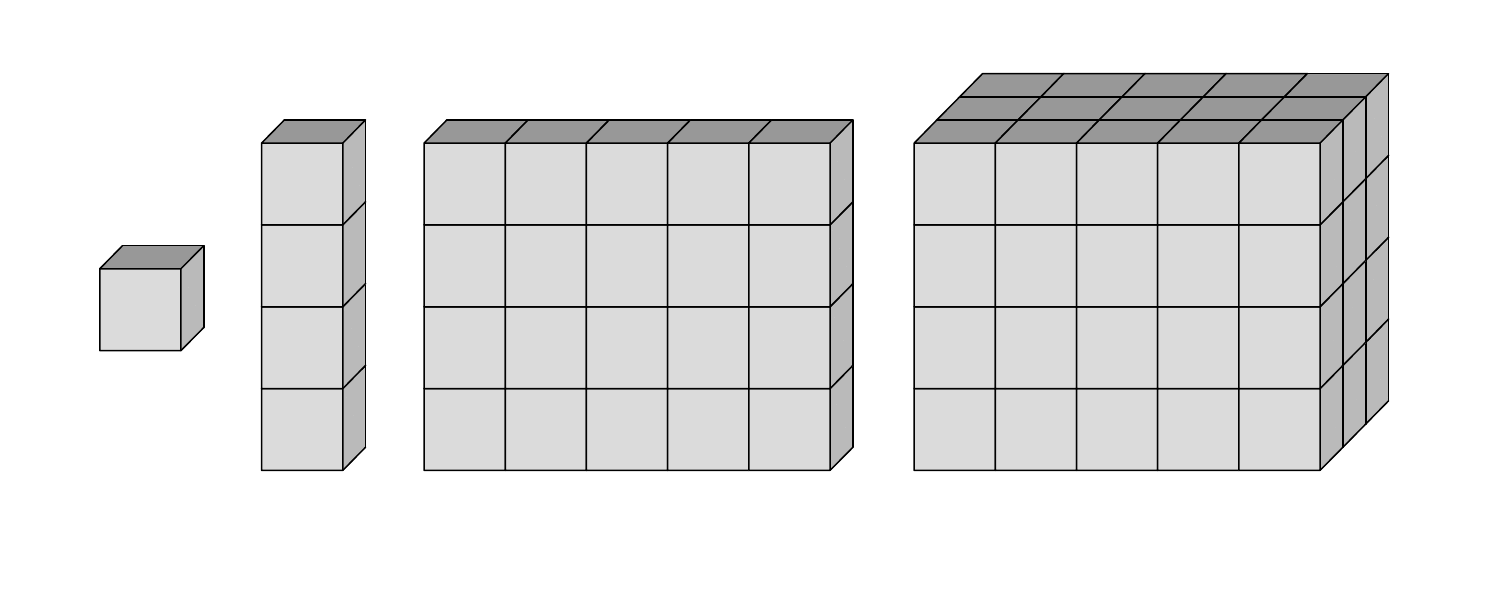}
	\caption{A scalar, $1^{\text{st}}$ order, $2^{\text{nd}}$ order, and $3^{\text{rd}}$ order tensors}\label{fig:tensors}
\end{figure}

While the two-dimensional panel data are represented by matrices, the multi-dimensional panel data lead to their higher-order counterparts, called tensors. A $d$-dimensional tensor is described by enumerating all the entries along the $d$ dimensions:
$\mathbf{Y}$ = $\left\{y_{i_1,i_2,\dots, i_d},\; 1\leq i_j\leq N_j,\; 1\leq j\leq d\right\};$
see Figure~\ref{fig:tensors}.

\smallskip

Tensor datasets are often characterized by complex dependencies between entries, so conventional econometric methods may not always be appropriate. For instance, the regional macroeconomic indicators are correlated with each other, as well as over space and time. \cite{babii2022tensor} consider the tensor factor model to capture such dependencies. For a tensor $\mathbf{Y}\in\R^{N\times J\times T}$, the model with $R$ latent factors is\footnote{Note that while one could consider a three-dimensional tensor as a collection of matrices and apply the standard factor model, this approach has significant limitations. It ignores the tensor structure and leads to the overparametrized model compared to the tensor factor approach.}
\begin{equation*}
	\mathbf{Y} = \sum_{r=1}^R\lambda_r\otimes \mu_r\otimes f_r + \mathbf{U},\qquad \E\mathbf{U}=0,
\end{equation*}
where $\mathbf{U}\in\R^{N\times J\times T}$ is an idiosyncratic noise tensor, $f_r\in\R^T$ are time series factors, $\lambda_r\in\R^N$ and $\mu_r\in\R^J$ are loadings in different dimensions, and we use the tensor product notation so that $(\lambda_r\otimes \mu_r\otimes f_r)_{i,j,t}=\lambda_{r,i}\mu_{r,j}f_{r,t}$. More generally, for a $d$-dimensional tensor $\mathbf{Y}\in\R^{N_1\times\dots\times N_d}$, the tensor factor model is
\begin{equation*}\label{eq:normalizedmodel}
	\mathbf{Y} = \sum_{r=1}^R\sigma_r\bigotimes_{j=1}^d m_{j,r} + \mathbf{U},\qquad \E\mathbf{U}=0,
\end{equation*}
where $m_{j,r}$ are the unit norm loadings/factors and $\sigma_r$ are the scale components.

\smallskip

\cite{babii2022tensor} study the PCA estimators for the tensor factor model based on tensor matricizations along each of its dimensions. The PCA estimators have a closed-form expression in contrast to the conventionally used alternating least-squares algorithm for tensor decomposition; see \cite{kolda2009tensor}. To describe the algorithm, we need to matricize tensors which is an operation generalizing the matrix vectorization:
\begin{example}
	Let $\mathbf{Y}$ be a $3\times 4\times 2$ dimensional tensor of the following two slices:
	\begin{equation*}
		\mathbf{Y}_1=\begin{bmatrix}
			1 & 4 & 7 & 10\\
			2 & 5 & 8 & 11\\
			3 & 6 & 9 & 12
		\end{bmatrix} \quad
		\mathbf{Y}_2=\begin{bmatrix}
			13 & 16 & 19 & 22\\
			14 & 17 & 20 & 23\\
			15 & 18 & 21 & 24
		\end{bmatrix}.
	\end{equation*}
	Then the mode-$1$, $2$ and $3$ matricizations of $\mathbf{Y}$ are respectively:
	\begin{equation*}
		\mathbf{Y}_{(1)} = \begin{bmatrix}
			1 & 4 & 7 & 10 & 13 & 16 & 19 & 22\\
			2 & 5 & 8 & 11 & 14 & 17 & 20 & 23\\
			3 & 6 & 9 & 12 & 15 & 18 & 21 & 24
		\end{bmatrix},\qquad \mathbf{Y}_{(2)} = \begin{bmatrix}
		1 & 2 & 3 & 13 & 14 & 15\\
		4 & 5 & 6 & 16 & 17 & 18\\
		7 & 8 & 9 & 19 & 20 & 21\\
		10 & 11 & 12 & 22 & 23 & 24
	\end{bmatrix},
	\end{equation*}
	\begin{equation*}
		\mathbf{Y}_{(3)} = \begin{bmatrix}
			1 & 2 & 3 & 4 & \cdots & 9 & 10 & 11 & 12\\
			13 & 14 & 15 & 16 & \cdots & 21 & 22 & 23 & 24
		\end{bmatrix}
	\end{equation*}
\end{example}

\noindent This leads to the following tensor PCA algorithm:
\begin{enumerate}
	\item Matricize the tensor $ \mathbf{Y} $ into matrices $\mathbf{Y}_{(1)}, \mathbf{Y}_{(2)}, \ldots, \mathbf{Y}_{(d)}$ along each of its dimensions.
	\item Estimate the unit norm factors and loadings as $(\hat m_{j,1},\dots,\hat m_{j,R})$ via PCA in each of the $d$ dimensions, i.e. the first $R$ eigenvectors of $N_j\times N_j$ matrix $\mathbf{Y}_{(j)}\mathbf{Y}_{(j)}^\top$.
	\item Estimate $(\hat\sigma^2_{r,j})_{r=1}^R$ as the $R$ largest eigenvalues of $\mathbf{Y}_{(j)}\mathbf{Y}_{(j)}^\top$.
\end{enumerate}

\smallskip

To determine the number of factors in a tensor factor model, \cite{babii2022tensor} consider the eigenvalue ratio test, cf.\ \cite{onatski2009testing}. They show that the null hypothesis that there are at most $k$ factors, we have for every matricization $j=1,2,\dots,d$,
\begin{equation*}
	S_j := \max_{k<r\leq K}\frac{\hat\sigma^2_{r,j} - \hat\sigma^2_{r+1,j}}{\hat\sigma^2_{r+1,j} - \hat\sigma^2_{r+2,j}} \xrightarrow{d}\max_{0<r\leq K-k}\frac{\xi_r - \xi_{r+1}}{\xi_{r+1} - \xi_{r+2}} =: Z,
\end{equation*}
where $(\xi_1,\dots,\xi_{K-k+2})$ follow the joint type-1 Tracy-Widom distribution; see \cite{karoui2003largest}. On the other hand, under the alternative hypothesis that the number of factors is $>k$ but $\leq K$, the statistics diverges to infinity. The testing procedure is
\begin{enumerate}
	\item Let $(Z_i)_{i=1}^m$ be $m$ independent random variables drawn from the same distribution as $Z$. To approximate the distribution of $(\xi_1,\xi_2,\dots)$, we can use the first eigenvalues of a symmetric $N_j\times N_j$ Gaussian matrix matrix with entries $\zeta_{i,j}\sim_{\rm i.i.d.} N(0,\tau_{i,j}),i\leq j$, where $\tau_{i,j}=1$ if $i<j$ and $\tau_{i,j}=2$ if $i=j$.
	\item Compute the p-value $p_j = 1-F_m(S_j)$ for each $1\leq j\leq d$, where $F_m(x)=\frac{1}{m}\sum_{i=1}^m\one_{Z_i\leq z}$.
	\item Combine the  p-values from the individual matricizations as $p_{\rm mean} = \frac{2}{d}\sum_{j=1}^dp_j$; see \cite{vovk2020combining}.
\end{enumerate}

\smallskip

There also exist tensor factor models related to the Tucker decomposition that allows for a different number of factors in different dimensions, see \cite{chen2022factor} and \cite{han2022rank}. While these models are more general they feature a larger number of parameters and may involve non-trivial identification issues. To the best of our knowledge, there are no formal statistical tests that can be used to decide which model is appropriate for a particular application.

\section{Conclusions}
Machine learning methods are attracting significant attention in economics and finance. The success of these methods stems from their ability to provide flexible regularized approximations to the theoretically optimal decision rules in data-rich environments. The empirical application of machine learning in economics and finance involves several methodological challenges.

\smallskip

In this survey, we cover some of the interesting recent developments that address these challenges. Many of the important topics were unfortunately omitted due to constraints. For example, we have barely touched the nonparametric ML methods such as boosting, random forests, support vectors machines, or neural networks; see   \cite{xudeep}, \cite{lahiri2022boosting},  \cite{medeiros2021forecasting}, \cite{bredahl2016forecasting}, \cite{rossi2015modeling}, \cite{bai2009boosting}, and references therein among many others.

\bigskip

\bibliographystyle{econometrica}
\bibliography{handbook-chapter-ml}

\end{document}